\documentclass[prb,twocolumn,aps,showpacs,fixfloats]{revtex4}
\usepackage{graphicx}
\usepackage{bm}
\usepackage{amsmath,amssymb}
\usepackage{subfigure}
\usepackage{float}
\usepackage{latexsym}
\usepackage{color}
\usepackage{enumerate}
\usepackage{pdfpages}
\usepackage{tikz}
\usepackage{hyperref}
\usepackage{graphicx}% complex graphics
\usepackage{bm}% bold math
\usepackage{amssymb} %math symbols
\usepackage{amsmath}
\usepackage{subfigure}

\usepackage{relsize}
\begin{document}

\title{Damping of the Franz-Keldysh oscillations in the presence of disorder}

\author{R. E. Putnam, Jr. and M. E. Raikh}

\affiliation{ Department of Physics and Astronomy, University of Utah, Salt Lake City, UT 84112}

\begin{abstract}
Franz-Keldysh oscillations of the optical absorption
in the presence of short-range disorder are studied
theoretically. The magnitude of the effect depends on the
relation between the mean-free path in a zero field and the
distance between the turning points in electric field.
Damping of the Franz-Keldysh oscillations by the disorder
develops at high absorption frequency.
Effect of damping is amplified by the fact that, that electron and hole
are most sensitive to the disorder near the
turning points. This is because, near the turning points, velocities of electron and
hole turn to zero.

\end{abstract}

\maketitle
\section{Introduction}
Application of electric field to an insulator modifies the shape of the absorption
spectrum in two respects: (i) below-threshold part of the spectrum develops a tail
which becomes more voluminous  upon increasing the field; (ii)  above-the-threshold part
of the spectrum acquires an oscillating component with a period of oscillations gradually
decreasing as the field is increased.
This effect predicted by Franz and Keldysh\cite{Franz,Keldysh}
more than 60 years ago was subsequently observed on a large number of bulk semiconductors,
see e.g. Ref.~\onlinecite{bulk}.
\begin{figure}
\includegraphics[width = 90mm]{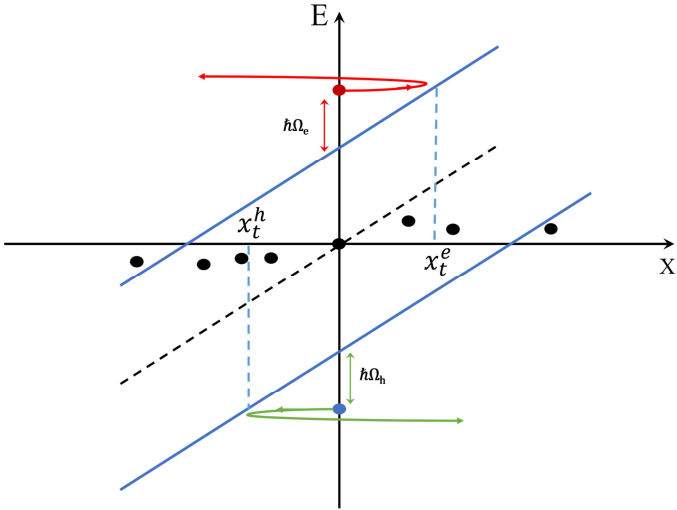}
\caption{(Color online) Schematic illustration of the above-the-gap electroabsorption in an
insulator. Initital kinetic energies of electron and hole are $\hbar\Omega_e$ and $\hbar\Omega_e$, respectively.
Franz-Keldysh oscillations reflect the interference of incident electron (hole) wave with a wave reflected at the corresponding turning point. Disorder, illustrated with dots, leads to the smearing of the interference pattern, and, consequently, to
the smearing of the oscillations. It is crucial that patterns for electron and hole
are smeared by scattering from {\em different} impurities.}
\label{figure1}
\end{figure}
%bulk semiconductors,\cite{bulk}
Later, theoretical and experimental studies were extended to quantum wells\cite{Chemla,Well} and quantum wires.\cite{Wire,Wire1}
Less trivial is that Franz-Keldysh oscillations were observed in certain polymers,\cite{polymer,polymer1}
and recently, in nanotubes,\cite{nanotube,nanotube1}
nanocrystals,\cite{nanocrystal} and perovskites.\cite{perovskite0,perovskite1,perovskite2}
Polymers and perovskites are known to be highly disordered materials with carrier mobility\cite{mobility} $\sim 10~\text{cm}^2\cdot \text{V}^{-1}\cdot \text{s}^{-1}$.
Then the question arises, how the inherent disorder in these materials affects the Franz-Keldysh
effect. Apparently, the effect of an electric field on the tail part of the absorption
spectrum is weakened by the disorder. This is because the disorder creates a tail of its own.
Effect of disorder on the above-the-threshold oscillations is of a different physical origin.
It is illustrated in Fig.~\ref{figure1}. Upon absorption of a photon, electron and hole
have kinetic energies $\hbar\Omega_e$ and $\hbar\Omega_h$, respectively. They get, subsequently,
reflected at the points $\pm x_t$, shown in the figure, and interfere with incident  waves. The
resulting density pattern resembles a standing wave. Presence of impurities "scrambles" the
standing-wave pattern, leading to the suppression of the Franz-Keldysh oscillations.
This suppression is studied theoretically in the present paper. Superficially, it is clear that,
in order to distinguish between the weak and strong disorder limits, one has to compare $x_t$ to the
mean free path, $l$. For a weak disorder one has $l\gg x_t$, so the oscillations are not affected.
However, there is a nontrivial aspect to this picture, namely, upon approaching the
corresponding turning points,
$\pm x_t$,
kinetic energies of electron and hole turn to zero. Thus, the local mean free paths decrease dramatically.
This contributes to the suppression of the interference, and thus, to the damping of the oscillations.

\section{Smearing of the Franz-Keldysh oscillations by the disorder}

Without the disorder, the oscillating part of the absorption coefficient of light with frequency, $\omega$, in the presence
of electric field, $F$, is given by

\begin{equation}
\label{Without}
A(\Omega)\propto \sin\left[c\left(\frac{\hbar\Omega}{E_0}\right)^{3/2}  \right],
\end{equation}
where $c$ is a constant,  and $\hbar\Omega=\hbar\omega -E_g$, where $E_g$ is a band-gap. For equal masses of electron and hole one has $c=\frac{2^{5/2}}{3}$.
Energy, $E_0$, in Eq. (\ref{Without}) is defined as
\begin{equation}
\label{E0}
E_0=\left(\frac{\hbar^2F^2}{m}   \right)^{1/3}.
\end{equation}
It has a meaning of the characteristic electron energy in the field, $F$.

Our main result is that, in the presence of disorder, the spectrum $A(\Omega)$ is multiplied by a damping factor $\exp\big[-{{\cal L}(\Omega)}   \big]$,
where the function ${\cal L}\left(\Omega\right)$ is defined as
\begin{equation}
\label{calL}
{\cal L}\left(\Omega\right)=\frac{m\gamma}{\hbar^2F}\ln\left(\frac{\hbar\Omega}{E_0}\right).
\end{equation}

Here $\gamma$ is the factor in the correlation relation,
\begin{equation}
\label{gamma}
\langle V(x)V(x')\rangle =\gamma \delta(x-x'),
\end{equation}
of the disorder potential, $V(x)$. Disorder leads to a finite scattering time. For electron with energy, $E$, the golden-rule calculation of the scattering time yields
\begin{equation}
\label{tau}
\frac{\hbar}{\tau(E)}=\frac{\gamma}{2}\left( \frac{2m}{\hbar^2E}  \right)^{1/2}.
\end{equation}
Expression in the square bracket is a 1D density of states.
Upon approaching the turning point, kinetic energy of electron decreases,
and, thus, the scattering time shortens. Decrease of velocity upon approaching the turning point leads to additional shortening of the mean free path.

\begin{equation}
\label{path}
l_E=\Big(\frac{2E}{m}\Big)^{1/2}\tau(E)=\frac{2\hbar^2}{m\gamma}E.
\end{equation}

Note, that oscillating absorption Eq. (\ref{Without})
corresponds to the large-argument asymptote of the square of the Airy function.
It applies for $\hbar\Omega \gg E_0$ when the argument of  sine is big.

Physical meaning of the prefactor $\frac{m\gamma}{\hbar^2F}$ in
Eq. (\ref{calL}) is the following. Energy scale $E_0$ comes from the
distance $x_F\sim \frac{E_0}{F}$. From Eq. (\ref{gamma}) it follows that
the r.m.s. random potential within the distance $x_F$ can be estimated as
$V(x_F)\sim \left(\frac{\gamma}{x_F}\right)^{1/2}$. Then the prefactor in Eq. (\ref{calL}) can be cast in the
form $\left(\frac{ V(x_F)}{E_0}    \right)^2$.

\section{General expression for the absorption coefficient}
The golden-rule expression for the absorption coefficient of light with frequency, $\omega$, is the following
\begin{align}
\label{GoldenRule}
&A(\omega)\nonumber\\
&\propto \sum_{\mu,\nu} \Big{\vert} \int d{\bf r}\psi_{\mu}^e({\bf r})\left(\psi_{\nu}^h({\bf r})\right)^{\ast}\Big{\vert}^2\delta\Big [\hbar\Omega_{\mu}^e+\hbar\Omega_{\nu}^h-\left(\hbar \omega -E_g   \right)     \Big],
\end{align}
where $\psi_{\mu}^e({\bf r})$, $\psi_{\nu}^h({\bf r})$ are the eigenfunctions of the initial and final states,
%electron and hole,
while $\hbar\Omega_{\mu}^e$ and $\hbar\Omega_{\nu}^h$ are the corresponding energies. Since the
eigenfunctions  carry information about the disorder, it is convenient to "decouple" the
$\delta$-function as follows
\begin{align}
\label{decouple}
&\delta \Big(\Omega_{\mu}^e+\Omega_{\nu}^h-\omega +\frac{E_g}{\hbar}    \Big)=
\int
d\Omega_1 \int d\Omega_2\nonumber\\
&\times \delta\left(\omega -\frac{E_g}{\hbar} -\Omega_1- \Omega_2  \right)
\delta\left(\Omega_1-\Omega_{\mu}^e  \right)\delta\left(\Omega_2-\Omega_{\nu}^h  \right),
\end{align}
and rewrite $A(\omega)$ in the form
\begin{align}
\label{GoldenRule1}
&A(\omega)=\int d\Omega_1\int d\Omega_2\delta\left(\Omega_1+\Omega_2+\frac{E_g}{\hbar}-\omega\right)                                    \nonumber\\
&\times \int d{\bf r}_1\int d{\bf r}_2\text{Im}G_e({\bf r}_1,{\bf r_2},\Omega_1)\text{Im}G_h({\bf r}_1,{\bf r_2},\Omega_2),
%=\!\int \!d\Omega_1\!\int \!d\Omega_2\delta\left(\Omega_1+\Omega_2+\frac{E_g}{\hbar}-\omega\right)
\end{align}
where the imaginary parts of the Green functions are defined as

\begin{align}
\label{G}
&\text{Im}G_e({\bf r}_1,{\bf r}_2,\Omega_1)=\sum_{\mu}\psi_{\mu}^e({\bf r}_1)
\left(\psi_{\mu}^e({\bf r}_2)\right)^{\ast}\delta\left(\Omega_{\mu}^e-\Omega_1   \right),\nonumber\\
&\text{Im}G_h({\bf r}_1,{\bf r}_2,\Omega_2)=\sum_{\nu}\left(\psi_{\nu}^h({\bf r}_1)\right)^{\ast}
\psi_{\nu}^h({\bf r}_2)\delta\left(\Omega_{\nu}^h-\Omega_2   \right).
\end{align}
In the presence of disorder, the product $\text{Im}G_e\text{Im}G_h$ in the  right-hand side
should be averaged over different configurations. It is crucial, that the disorder-induced smearing of the
interference patterns for electrons and holes is dominated by {\em different} impurities. This
is illustrated in Fig.~\ref{figure1}. As a result, the disorder averaging can be performed {\em independently}.

\section{Semiclassical calculation of the absorption damping in 1D}

\begin{figure}
\includegraphics[width = 90mm]{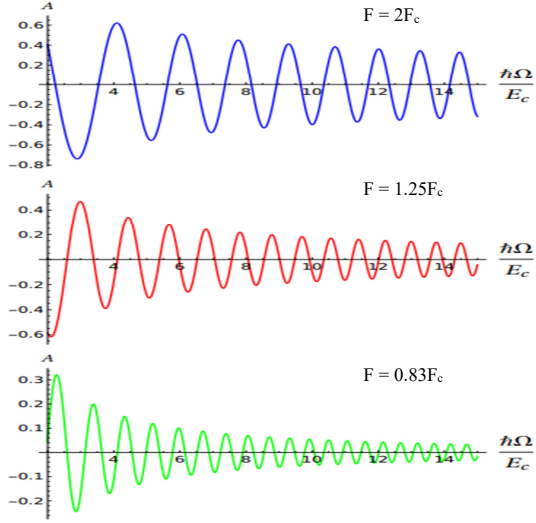}
\caption{(Color online) Evolution of Franz-Keldysh oscillations with applied electric field is
  plotted from Eqs. (\ref{Without}) and (\ref{calL}) for three values of electric field, $F$,
measured in the units of $ F_c$  given by Eq. (\ref{Fc}). Energy $\hbar\Omega$ is measured
in the units of $E_c=\left( \frac{m\gamma^2}{2\hbar^2}  \right)^{1/3}$. }
\label{figure2}
\end{figure}

Franz-Keldysh oscillations emerge in semiclassical regime, when electron and hole states
can be described by local wave vectors
\begin{equation}
\label{k_e}
k_e(x)=\left(\frac{2m}{\hbar^2}\right)^{1/2}\left(\hbar\Omega_1-Fx\right)^{1/2},
\end{equation}

\begin{equation}
\label{k_e}
k_h(x)=\left(\frac{2m}{\hbar^2}\right)^{1/2}\left(\hbar\Omega_2+Fx\right)^{1/2}.
\end{equation}
In accordance with Fig. \ref{figure1}, $k_e(x)$ turns to zero at the right turning point, while $k_h(x)$ turns to zero
at the left turning point.
Electron and hole Green functions, which enter the light absorption Eq. (\ref{GoldenRule1})
can be expressed in terms of $k_e(x)$ and $k_h(x)$ as follows
\begin{align}
\label{ImGe}
&\text{Im}
G_e^{(0)}(x_1,x_2,\Omega_1)=
\frac{\cos\left(\Phi_e(x_1)-\frac{\pi}{4}\right)\cos\left(\Phi_e(x_2)-\frac{\pi}{4}\right)}
{\left(k_e(x_1)k_e(x_2)   \right)^{1/2}},\nonumber\\
&\text{Im}G_h^{(0)}(x_1,x_2,\Omega_2)=
\frac{\cos\left(\Psi_h(x_1)+\frac{\pi}{4}\right)\cos\left(\Psi_h(x_2)+\frac{\pi}{4}\right)}
{\left(k_h(x_1)k_h(x_2)   \right)^{1/2}}.
\end{align}
where the semiclassical actions $\Phi_e$ and $\Phi_h$ are defined as
\begin{equation}
\label{Phi}
\Phi_e(x_1)=\int\limits_{x_1}^{\hbar\Omega_1/F}dx'k_e(x'),~
\Psi_h(x_1)=\int\limits_{-\hbar\Omega_2/F}^{x_1}dx'k_h(x').
\end{equation}
Expressions Eq. (\ref{ImGe}) are semiclassical and apply away from the turning points:
\begin{equation}
\label{condition}
\Big( \frac{\hbar\Omega_1}{F}-x_1\Big)\gg \Big( \frac{\hbar^2}{mF} \Big)^{1/3}=x_F,~
\Big(x_1+\frac{\hbar\Omega_2}{F}\Big)\gg x_F.
\end{equation}

Our goal is to incorporate disorder into the Green functions. For this purpose we
transform the products of the cosines into the sums
\begin{align}
\label{G1}
&\text{Im}
G_e^{(0)}(x_1,x_2,\Omega_1)\nonumber\\
&=\frac{\cos\left(\Phi_e(x_1)-\Phi_e(x_2)\right)+\cos\left(\Phi_e(x_1)+\Phi_e(x_2)-\frac{\pi}{2}\right)}
{\left(k_e(x_1)k_e(x_2)   \right)^{1/2}},
\end{align}

\begin{align}
\label{G2}
&\text{Im}G_h^{(0)}(x_1,x_2,\Omega_2)\nonumber\\
&=\frac{\cos\left(\Psi_h(x_1)-\Psi_h(x_2)\right)+\cos\left(\Psi_h(x_1)+\Psi_h(x_2)+\frac{\pi}{2}\right)}
{\left(k_h(x_1)k_h(x_2)   \right)^{1/2}}.
\end{align}
Physically, the first term in Eq. (\ref{G1}) describes the propagation of electron from $x_1$
to $x_2$, while the second term describes the motion of electron from $x_1$ to the turning point,
$x=x_t$ followed by reflection from the electrostatic barrier and return to $x_2$.
It is shown in the Appendix in great detail that disorder averaging of the first term amounts
to multiplying this term by $\exp(-|x_2-x_1|/l)$ when the mean free path, $l$, does not depend on $x$.
Since, in electric field, kinetic energy of electron is a function of position, and  $l$ is a function of energy,
we conclude that the mean free path is a function of position.
% and thus, in the presence of electric field, it is a function of coordinate.
Then the natural generalization of the damping factor is
$\exp\Big(-\int\limits_{x_1}^{x_2}\frac{dx}{l(x)}\Big)$.
Substituting the $x$-dependent kinetic energy into Eq. (\ref{path}),
we get the following expressions for the mean free paths of electrons and holes
\begin{equation}
\label{L}
l_1(x)^{-1}=\frac{m\gamma}{2\hbar^2(\hbar\Omega_1-Fx)},~l_2(x)^{-1}=\frac{m\gamma}{2\hbar^2(\hbar\Omega_2+Fx)}.
\end{equation}
Thus, incorporation of disorder into the Green functions $\text{Im}
G_e^{(0)}(x_1,x_2)$ and $\text{Im}G_h^{(0)}(x_1,x_2)$ amounts to ascribing to
each cosine in Eqs. (\ref{G1} and (\ref{G2}) its corresponding damping factor.

Turning to the product $\text{Im}G_e^{(0)}\text{Im}G_h^{(0)}$ we notice that the
piece of this product that captures the Franz-Keldysh oscillations involves both the
reflection of electron from electrostatic barrier at $x=x_t$ and reflection of hole from
electrostatic barrier at $x=-x_t$. This process is encoded into the product
%Thus, the piece of the product, $\text{Im}G_e^{(0)}\text{Im}G_h^{(0)}$, responsible
%for the Franz-Keldysh oscillations,
\begin{align}
\label{cosine}
&\cos\left(\Phi_e(x_1)+\Phi_e(x_2)-\frac{\pi}{2}\right)
\cos\left(\Psi_h(x_1)+\Psi_h(x_2)+\frac{\pi}{2}\right)\nonumber\\
&=\frac{1}{2}\Bigg[ -\cos\Big( \Phi_e(x_1)- \Psi_h(x_1)+ \Phi_e(x_2)- \Psi_h(x_2)\Big) \nonumber\\
& +\cos\Big(\Phi_e(x_1)+\Psi_h(x_1)+\Phi_e(x_2)+ \Psi_h(x_2)\Big)\Bigg].
\end{align}
It is the second cosine that captures two reflections. The damping factor
corresponding to this cosine has the form $\exp\left(-{\cal L}   \right)$,
where ${\cal L}$ is given by
\begin{equation}
\label{calL1}
{\cal L}=\int\limits_{-\hbar\Omega_2/F}^{x_1}\frac{dx}{l_h(x)}+\int\limits_{x_1}^{\hbar\Omega_1/F}
\frac{dx}{l_e(x)}+\int\limits_{-\hbar\Omega_2/F}^{x_2}\frac{dx}{l_h(x)}+\int\limits_{x_2}^{\hbar\Omega_1/F}
\frac{dx}{l_e(x)}.
\end{equation}
The remaining task is to substitute Eqs. (\ref{cosine}) and (\ref{calL1}) into Eq. (\ref{GoldenRule1})
and perform the integration over $\Omega_1$, $\Omega_2$, $x_1$, $x_2$. Integration is determined
by rapidly changing cosine Eq. (\ref{cosine}). Concerning the damping factor, it is sufficient to
substitute $\Omega_1=\Omega_2=\frac{E_g-\hbar\omega}{2\hbar}$ and set $x_1=x_2=0$. Integrals in
Eq. (\ref{calL1}) diverge logarithmically. A natural cutoff of the logarithms is $x_F$,
see Eq. (\ref{condition}).
Then Eq. (\ref{calL1}) reduces to our main result Eq. (\ref{calL}).

\section{Concluding remarks}
\begin{figure}
\includegraphics[width = 90mm]{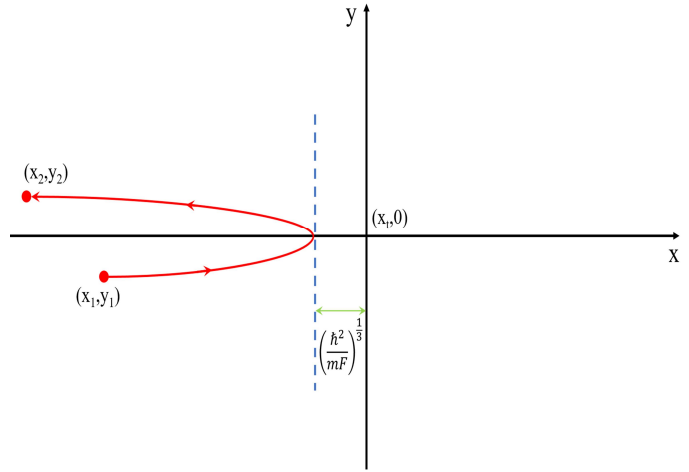}
\caption{(Color online) Illustration of the Franz-Keldysh oscillations in 2D. Semiclassical trajectory of electron from the
point $x_1,y_1$  towards the electrostatic barrier is close to the trajectory from the barrier to the point $x_2,y_2$. Electron, moving
along the semiclassical trajectory is scattered by the disorder, which ``scrambles" the interference pattern responsible for the oscillations. }
\label{figure3}
\end{figure}

In traditional semiconductor materials the shape of the
Franz-Keldysh oscillations is quite robust.\cite{Sipe}
Previous theoretical studies\cite{exciton1,Aspnes,Dow,Merkulov,Dow1,Aronov}
were focused on the effect of electron-hole attraction on electroabsorption.
Numerical calculations\cite{Sipe} indicate that the excitonic effect does not alter
the shape of the oscillations. Rather,  a smooth background on which
Franz-Keldysh oscillations develop becomes more pronounced due to the electron-hole
attraction.\cite{Sipe} Effect of disorder on the oscillations is illustrated in Fig. \ref{figure2},
where the oscillations are plotted from Eqs. (\ref{Without}) and (\ref{calL})
for a given disorder and three values of electric field. It is seen that, while
disorder damps the oscillations, this damping becomes less effective upon increasing, $F$.
Physically, upon increasing
$F$, the travel distance of electron and hole from
the respective turning points to the point of their meeting becomes shorter.
To estimate quantitatively the critical field, $F_c$, above which the effect of disorder becomes
negligible, we express $F_c$ in terms  of the width, $E_c$,  of the tail in the density of states,
which disorder creates {\em in the absence of external field}.
The width, $E_c$, can be estimated by equating the uncertainty, $\frac{\hbar}{\tau(E_c)}$, given
by Eq.~(\ref{tau}) to $E_c$.
This yields $E_c=\left(\frac{\gamma}{2}\right)^{2/3}\left(\frac{2m}{\hbar^2}\right)^{1/3}$.
Now the parameter $\gamma$ can be expressed via the observable $E_c$. Then, equating to one the
prefactor, $\frac{m\gamma}{\hbar^2F}$,
in front of logarithm in Eq. (\ref{calL}) yields the value
\begin{equation}
\label{Fc}
F_c=\frac{\left(mE_c^3\right)^{1/2}}{\hbar}.
\end{equation}
For example, setting $m$ equal to the free electron mass, for disordered system with a tail
$E_c=0.1~\text{eV}$, we conclude that Franz-Keldysh oscillations are smeared when the applied field
is smaller than $1.5~\text{kV}\cdot \text{cm}^{-1}$. Note that, the tail, $E_c$,
can be related to the mean free path
$l_c=\left(\frac{\hbar^2}{mE_c}\right)^{1/2}\approx 10~\mathring{A}$.
The above values are consistent with those reported for polymers.\cite{polymer,polymer1}
Overall, quantitative comparison of theoretical results with numerous experimental data
is complicated by the fact that experimental papers do not list the values of electric field,
but rather the voltage drop on the sample. Qualitatively though, experimental data on the Franz-Keldysh
oscillations are in agreement with our prediction:
Decrease of the oscillation period with their number
is weak, while the visibility of high-order oscillations increases with increasing, $F$.
To account for this effect, the authors of Refs. \onlinecite{polymer}, \onlinecite{polymer1} adopted a phenomenological
approach: in calculation of absorption spectrum the electron energy, $E$,
is replaced by $E+i\Gamma$, where parameter $\Gamma$ emulates the disorder. It is used as a fitting parameter.
In this regard, within our theory, the role of
``coherence length" in Refs. \onlinecite{polymer}, \onlinecite{polymer1} is played by the mean free path due to impurity scattering and {\em depends strongly on}, $F$. 
We have restricted consideration to the 1D case. The result in 2D
does not differ qualitatively. This is because the electron travel path towards the turning point lies close to
the travel path of the reflected electron. This is illustrated in Fig. \ref{figure3}. Quantitatively, the  difference between the 2D and 1D cases is that the density of states in 2D
is constant, so  that the mean free path grows with energy slower, as $E^{1/2}$  rather than as $E$.

\appendix

\section{Average Green's function near a wall}

In the presence of a wall at $y=0$, the Green's function of a free electron
in the half-space $y>0$ has a form
\begin{equation}
\label{bare}
G^{(0)}\left(E,y,y'\right)=\frac{im}{\hbar^2k_{E}}
\Big[e^{-ik_E|y-y'|}-e^{-ik_E(y+y')} \Big],
\end{equation}
where $k_E=\left(\frac{2mE}{\hbar^2} \right)^{1/2}$ is a wave vector.
The function $G^{(0)}$ satisfies the boundary conditions
$G^{(0)}\left(E,0,y'\right)=G^{(0)}\left(E,y,0\right)=0$.

\begin{figure}
\includegraphics[width = 90mm]{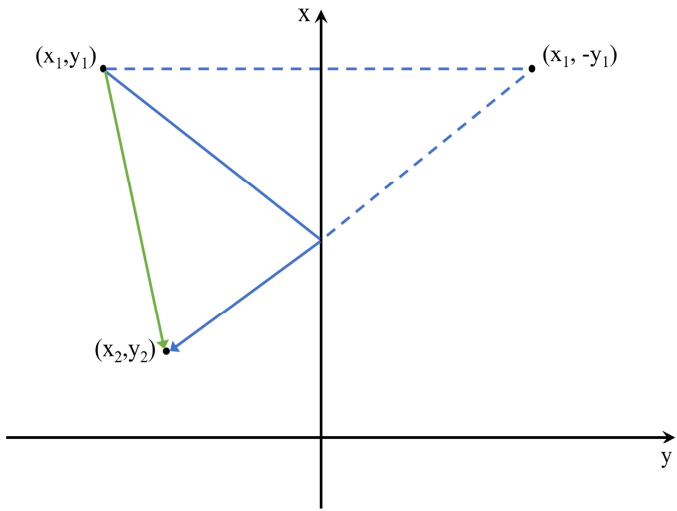}
\caption{(Color online) Illustration of the 2D Green's function in the presence of a wall at $y=0$.
There are two classical paths between the points $(x_1,y_1)$ and $(x_2,y_2)$. One path is along the
straight line, while the other path corresponds to the shortest classical trajectory reflected from the wall.}
\label{figure4}
\end{figure}

With a random disorder potential, $V(y)$, the Green's function
satisfies the Dyson equation
\begin{equation}
\label{Dyson}
G(y_1,y_2)=G^{(0)}(y_1,y_2)+\int\limits_0^{\infty}dy G^{(0)}(y_1,y)V(y)G(y,y_2).
\end{equation}
This equation is exact. It can be cast into a different form upon
substituting the right-hand side into the integrand
\begin{align}
\label{Dyson'}
&G(y_1,y_2)=G^{(0)}(y_1,y_2)+\int\limits_0^{\infty}dy G^{(0)}(y_1,y)V(y)G^{(0)}(y,y_2)\nonumber\\
&+\int\limits_0^{\infty}dy\int\limits_0^{\infty}dy' G^{(0)}(y_1,y)V(y)G^{(0)}(y,y')V(y')
G(y',y_2).
\end{align}
A crucial step which allows to get a closed equation for the average Green's function is decoupling
of averaging in the integrand of Eq. (\ref{Dyson'}). Assuming that the potential $V(y)$ is short-ranged
\begin{equation}
\label{correlation}
\langle V(y)V(y')\rangle =\gamma\delta(y-y'),
\end{equation}
and performing the averaging, we arrive to the closed RPA equation
\begin{align}
\label{Dyson''}
&\langle G(y_1,y_2)\rangle = G^{(0)}(y_1,y_2)\nonumber\\
&+\gamma \int\limits_0^{\infty}dy G^{(0)}(y_1,y)G^{(0)}(y,y)\langle G(y,y_2)   \rangle.
\end{align}

In the absence of the wall, the bare Green's function is equal to $G^{(0)}=\frac{im}{\hbar^2k_{E}}
e^{-ik_E|y-y'|}$ and the lower limit in the integral Eq. (\ref{Dyson''})
should be set to $-\infty$.
Then the average Green's function depends only on the difference $(y_1-y_2)$. This allows to solve Eq. (\ref{Dyson''})  with the help of the Fourier transform.
As a result, the disorder leads to a factor  $\exp\big(-\frac{|y_1-y_2|}{l}\big)$  in the average Green's function.
Here $l$ is the mean free path which can be expressed via the disorder strength as follows
\begin{equation}
\label{l}
l=\frac{\hbar^4k_E^2}{m^2\gamma}.
\end{equation}
The above result applies when the disorder is weak, namely, for $k_El \gg 1$.

In the presence of the wall, the translational symmetry is violated. Taking the Fourier transform is
not permissible.  Yet, the form of the disorder-modified average Green's function can be established
from the following reasoning. The travel distance between the points $y_1$ and $y_2$ is either $|y_1-y_2|$, when the
particle travels directly, or $(y_1+y_2)$, when the travel involves the reflection from the wall.
This suggests the following form

\begin{align}
\label{bare1}
&\langle G\left(E,y,y'\right)\rangle\nonumber\\
&=\frac{im}{\hbar^2k_{E}}
\Big[e^{-ik_E|y-y'|}e^{-\frac{|y-y'|}{l}}-e^{-ik_E(y+y')}e^{-\frac{(y+y')}{l}} \Big].
\end{align}
of the average Green's function.

%\begin{figure}
%\includegraphics[width = 90mm]{KeldyshCartoon3.pdf}
%\caption{(Color online) Illustration of the Franz-Keldysh oscillations in 2D. Semiclassical trajectory of electron from the
%point $x_1,y_1$  towards the electrostatic barrier is close to the trajectory from the barrier to the point $x_2,y_2$. Electron, moving
%along the semiclassical trajectory is scattered by the disorder, which ``scrambles" the interference pattern responsible for the oscillations. }
%\label{figure3}
%\end{figure}

A nontrivial question is a minimal distance at which Eq.~(\ref{bare1}) applies.
We will show that the condition of applicability is $y_1,y_2 \gg \frac{1}{k_E}$.

To solve the equation Eq. (\ref{Dyson''}) approximately, we make a substitution
\begin{align}
\label{substitute}
&\langle G(y,y')\rangle\nonumber\\
&=\frac{im}{\hbar^2k_E}\Big[R_1(y,y')e^{-ik_E|y-y'|}-R_2(y,y')e^{-ik_E(y+y')}\Big],
\end{align}
and assume that $R_1$ and $R_2$ are  slow functions on the scale  $k_E^{-1}$.
Substituting the form Eq. (\ref{substitute}) into Eq. (\ref{Dyson''}), we cast
it into the form
\begin{widetext}
\begin{align}
\label{substitution1}
&\Big( R_1(y,y')-1  \Big)e^{-ik_E|y-y'|}-\Big( R_2(y,y')-1  \Big)e^{-ik_E(y+y')}\nonumber\\
&=\gamma \left(\frac{im}{\hbar^2k_E}  \right)^2\int\limits_0^{\infty}dy_1\Big(e^{-ik_E|y-y_1|}-e^{-ik_E(y+y_1)} \Big)\left(1-e^{-2ik_Ey_1}   \right) \Big(R_1(y_1,y')e^{-ik_E|y_1-y'|}-R_2(y_1,y')e^{-ik_E(y_1+y')}\Big).
\end{align}
\end{widetext}
Without the loss of generality we assume that $y'>y$.
The key step is dividing the integral in the right-hand-side
of Eq. (\ref{substitution1}) into three domains: (i) $0<y_1<y$,
(ii) $y<y_1<y'$, and (iii) $y_1>y'$. Consider first the domain (i). In this domain, we have
$|y-y_1|=y-y_1$, $|y_1-y'|=y'-y_1$. Out of six
terms in the integrand only two terms,
$-e^{-ik_E|y-y_1|}R_2(y_1,y')e^{-ik_E(y_1+y')}=-e^{-ik_E(y+y')}R_2(y_1,y')$,
and $-e^{-ik_Ey}R_1(y_1,y')e^{-ik_Ey'}=-e^{-ik_E(y+y')}R_1(y_1,y')$
do not oscillate with
$y_1$, since the functions  $R_1$ and $R_2$ are slow. All other terms, e.g.
$e^{-ik_E|y-y_1|}R_1(y_1,y')e^{-ik_E|y_1-y'|}=e^{-ik_E(y+y')}R_1(y_1,y')e^{2ik_Ey_1}$,
contain rapidly oscillating exponential factors. These factors suppress the
integral over the domain (i) down to $k_E^{-1}$. Similar strategy is applied
to the domains (ii) and (iii).  It can be checked that in the domain (iii) there are no
slow terms  in the integrand. In the domain (ii) there are two slow
terms in the integrand, namely $R_1(y_1,y')e^{-ik(y'-y)}$ and $R_1(y_1,y')e^{ik(y'-y)}$.
Collecting the terms with slow integrand and equating coefficients in front of
$e^{-ik_E|y-y'|}$ and $e^{-ik_E(y+y')}$ yields the system of equations

\begin{equation}
\label{R1}
R_1(y,y')=1-\gamma\left(\frac{im}{\hbar^2k_E}   \right)^2\int\limits_{y}^{y'}dy_1R_1(y_1,y').
\end{equation}

\begin{align}
\label{R2}
&R_2(y,y')=1+\gamma\left(\frac{im}{\hbar^2k_E}   \right)^2\int\limits_{0}^{y'}dy_1R_1(y_1,y') \nonumber\\ &+\gamma\left(\frac{im}{\hbar^2k_E}   \right)^2\int\limits_{0}^{y}dy_1R_2(y_1,y').
\end{align}
Exponential decay of $R_1$  follows from the fact that the
derivative $\frac{dR_1}{dy}$ is proportional to n$R_1$. It is easy to check that $R_1=\exp(-\frac{y'-y}{l})$
with $l$ defined by Eq. (\ref{l}) is the solution of Eq. (\ref{R1}).
Equivalently, the solution of Eq. (\ref{R2}) is $R_2=\exp(-\frac{y'+y}{l}   )$.
Note that in deriving the system Eqs. (\ref{R1}), (\ref{R2}) we assumed that $k_Ey\gg 1$, $k_Ey'\gg 1$,
when the terms with rapidly oscillating integrands can be neglected.

It is straightforward to extend Eq. (\ref{bare1}) to the 2D case. The Green function
is dominated by two classical paths. The first path with a length $\rho_1=\left[(x_1-x_2)^2+(y_1-y_2)^2\right]^{1/2}$
is along the straight line, while the second path involves the reflection from the wall. As illustrated
in Fig. \ref{figure4}, the minimal length of the path of this type  is $\rho_2=\left[(x_1-x_2)^2+(y_1+y_2)^2\right]^{1/2}$.
Within the numerical factor, the 2D Green's function has a form
\begin{align}
\label{2D}
&\langle G(E,x_1,y_1,x_2,y_2)\rangle \nonumber\\
&=\frac{m}{\left(2\pi\right)^{1/2}\hbar^{5/2}}\Bigg[\frac{\exp\left({-ik_E\rho_1-\frac{\rho_1}{l}}\right)}{(k_E\rho_1)^{1/2}}
-\frac{\exp\left(-{ik_E\rho_2-\frac{\rho_2}{l}}\right)}{(k_E\rho_2)^{1/2}}  \Bigg].
\end{align}

\section{Acknowledgements}

\vspace{2mm}

The work was supported by the Department of Energy,
Office of Basic Energy Sciences, Grant No.  DE- FG02-
06ER46313. We are grateful to Kameron Hansen, a graduate student' from Chemistry
Department, for picking our interest in electroabsorption in perovskites.


\vspace{10mm}

\begin{thebibliography}{30}



\bibitem{Franz}
W. Franz,
``Einfluss eines  elektrischen  Feldes  auf  eine  optische
Absorptionskante," Zeitschrift f{\"u}r Naturforschung {\bf 13a},
484 (1958).


\bibitem{Keldysh} L. V. Keldysh,
``Behaviour of Non-Metallic Crystals in Strong Electric Fields,"
Sov. Phys. JETP {\bf 6}, 763 (1958).

\bibitem{bulk}
T. E. Van Eck, L. M. Walpita, W. S. C. Chang, and H. H. Wieder,
``Franz-Keldysh
electrorefraction and electroabsorption in bulk InP and GaAs,"
Appl. Phys. Lett. {\bf 48}, 451 (1986).


\bibitem{Chemla}
D. A.  Miller, D. S. Chemla, and  S. Schmitt-Rink,
``Relation between Electroabsorption in Bulk Semiconductors and in Quantum Wells:
The Quantum-Confined Franz-Keldysh Effect,"
Phys. Rev. B {\bf 33}, 6976 (1986).

\bibitem{Well}
P. J. Hughes, B. L. Weiss, and J. Hosea,
``Analysis of Franz$-$Keldysh oscillations in photoreflectance
spectra of a
AlGaAs-GaAs
single-quantum well structure,"
Journ. of Appl. Phys. {\bf 77}, 6472 (1995).

\bibitem{Wire} T. Y. Zhang and W. Zhao, ``Franz-Keldysh effect and
dynamical Franz-Keldysh effect of cylindrical quantum wires,"
Phys. Rev. B {\bf 73}, 245337 (2006).

\bibitem{Wire1}
D. Li, J. Zhang, Q. Zhang, and Q. Xiong,
``Electric$-$Field$-$Dependent Photoconductivity in CdS Nanowires and Nanobelts:
Exciton Ionization, Franz-Keldysh, and Stark Effects,"
Nano Lett. {\bf 12}, 2993 (2012).

\bibitem{polymer}
A. Horvath, G. Weiser, C. Lapersonne-Meyer, M. Schott, and S. Spagnoli,
``Wannier excitons and Franz-Keldysh effect of polydiacetylene chains diluted
in their single crystal monomer matrix,"
Phys. Rev. B {\bf 53}, 13507 (1996).

\bibitem{polymer1}
G. Weiser, L. Legrand, T. Barisien, A. Choueiry, M. Schott, and S. Dutremez,
``Stark effect and Franz-Keldysh effect of a quantum wire realized by conjugated
polymer chains of a diacetylene 3NPh2,"
Phys. Rev. B {\bf 81}, 125209 (2010).

\bibitem{nanotube}
V. Perebeinos and P. Avouris,
``Exciton Ionization, Franz-Keldysh and Stark Effects in Carbon Nanotubes," Nano Lett. {\bf 7}, 609 (2007).


\bibitem{nanotube1}
M.-H. Ham, B.-S. Kong, W.-J. Kim, H.-T. Jung, and
M. S. Strano, ``Unusually Large Franz-Keldysh
Oscillations at Ultraviolet Wavelengths
in Single-Walled Carbon Nanotubes,"
Phys. Rev. Lett. {\bf 102}, 047402 (2009).

\bibitem{nanocrystal} N. V. Tepliakov,  M. Yu. Leonov,  A. V. Baranov,
A. V. Fedorov, and I. D. Rukhlenko, ``Quantum theory of electroabsorption in
semiconductor nanocrystals," Opt. Express  {\bf 24},  A52 (2016).




\bibitem{perovskite0}
E. Amerling, S. Baniya,
E. Lafalce, C. Zhang,
Z. V. Vardeny, and L. Whittaker-Brooks,
``Electroabsorption Spectroscopy Studies of (C4H9NH3)2PbI4
Organic$-$Inorganic Hybrid Perovskite Multiple Quantum Wells,"
J. Phys. Chem. Lett. {\bf 8}, 4557  (2017).

\bibitem{perovskite1}
 S. Xia, Z. Wang, Y. Ren, Z. Gu, and   Y. Wang,
 ``Unusual electric field-induced optical behaviors in cesium lead bromide
 perovskites," Appl. Phys. Lett. {\bf 115}, 201101 (2019).

\bibitem{perovskite2}
G. Walters, M. Wei, O. Voznyy, R. Quintero-Bermudez, A. Kiani,
D.-M. Smilgies, R. Munir,
A. Amassian, S. Hoogland, and E. Sargent,
``The quantum-confined Stark effect in layered
hybrid perovskites mediated by orientational
polarizability of confined dipole,"
Nat. Commun. {\bf 9}, 4214 (2018).

%\bibitem{mobility}
%S. M. Kwon, J. K. Won, J.-W. Jo, J. Kim, H.-J. Kim, H.-I. Kwon,
%J. Kim, S. Ahn, Y.-H. Kim, M.-J. Lee, H.-ik Lee, T. J. Marks,
%M.-G. Kim, and S. K. Park,
%``High-performance and scalable metal-chalcogenide semiconductors and
%devices via chalco-gel routes," Sci. Advances {\bf  4},  eaap9104 (2018).
%hybrid perovskites

\bibitem{mobility}
C. Motta, F. El-Mellouhi, and S. Sanvito,
``Charge carrier mobility in hybrid halide perovskites,"
Sci. Rep. {\bf 5}, 12746 (2015).


\bibitem{Sipe}
F. Duque-Gomez and J. E. Sipe,
``The Franz-Keldysh effect revisited: Electroabsorption
 interband coupling and excitonic," Journ. of Phys. Chem. Sol. {\bf 76}, 138 (2015).

 \bibitem{exciton1} H. I. Ralph, ``On the theory of Franz-Keldysh effect,"
 J. Phys. C: Solid State Phys. {\bf 1}, 378 (1968).




\bibitem{Aspnes}
D. E. Aspnes,
``Electric-Field Effects on Optical Absorption near Thresholds in Solids,"
Phys. Rev. {\bf 147}, 554, (1966).

\bibitem{Dow} J. D. Dow and D. Redfield,
``Electroabsorption in Semiconductors: The Excitonic Absorption Edge,"
Phys. Rev. B {\bf 1}, 3358 (1970).


\bibitem{Merkulov}
I. A. Merkulov, ``Influence of exciton effect on electroabsorption in
semiconductors," Zh. Eksp. Teor. Fiz. {\bf 66}, 2314   (1974).
 [Sov.
Phys. JETP {\bf 39}, 1140 (1974)].


\bibitem{Dow1}
 F. L. Lederrnan and J. D. Dow,
 ``Theory of electroabsorption by anisotropic and layered semiconductors. I. Two-dimensional
excitons in a uniform electric field," Phys. Rev. B {\bf 13}, 1633 (1976).


\bibitem{Aronov}
A. G. Aronov and A. S. Ioselevich,
``Effect of electric field on exciton absorption,"
Zh. Eksp. Teor. Fiz. {\bf 74}, 1043 (1978). [Sov. Phys. JETP {\bf 47}, 548  (1978)].




\end{thebibliography}
\end{document}